\documentclass[12pt,,letterpaper]{JHEP3}
\usepackage{graphics}
\usepackage{epsfig}
\title{Consistency Conditions on S-Matrix of Spin 1 Massless Particles}

\author{Song He\\
School of Physics, Peking University, Beijing, 100871, China\\
\email{hesong@pku.edu.cn}}
\author{Hongbao Zhang\\
Perimeter Institute for Theoretical Physics,
Waterloo, Ontario N2L 2Y5, Canada\\
Department of Astronomy, Beijing Normal University,
Beijing, 100875, China\\
\email{hzhang@perimeterinstitute.ca}}

\abstract{Motivated by new techniques in the computation of
scattering amplitudes of massless particles in four dimensions, like
BCFW recursion relations, the question that how much structure of
the S-matrix can be determined from purely S-matrix arguments has
received new attention. The BCFW recursion relations for massless
particles of spin 1 and 2 imply that the whole tree-level S-matrix
can be determined in terms of three-particle amplitudes evaluated at
complex momenta. However, the known proofs of the validity of the
relations rely on the Lagrangian of the theory, either by using
Feynman diagrams explicitly or by studying the effective theory at
large complex momenta. This means that a purely S-matrix proof of
the relations is still missing. The aim of this paper is to provide
such a proof for spin 1 particles by extending the four-particle
test introduced by Benincasa and Cachazo in~\cite{BC} to arbitrary
numbers of particles. We show how $n$-particle tests imply that the
rational function built from the BCFW recursion relations possesses
all the correct factorization channels including holomorphic and
anti-holomorphic collinear limits, which thus produces the correct
S-matrix of the theory.}

\begin{document}
\newpage
\tableofcontents
\newpage

\section{Introduction}
As an alternative to Feynman diagram analysis, BCFW construction is
a powerful tool for constructing tree-level amplitudes in terms of
sub-amplitudes with fewer external particles~\cite{BCF,BCFW}. The
class of theories whose amplitudes can be completely determined by
such a BCFW construction from lower amplitudes are called
constructible~\cite{BC}, and have been proven to include Yang-Mills
theory, General Relativity and more general two derivative theories
such as QCD, $\mathcal{N}=4$ SYM theory and $\mathcal{N}=8$
Supergravity~\cite{BCFW,BBST,CS,BBC,H,AK,Ch,ACK,DH,BEF}. However,
all of these proofs rely heavily on Lagrangian of the theory, either
by using Feynman diagrams explicitly~\cite{BCFW,BBST,CS,BBC,H}, or
by studying the effective theory at large complex
momenta~\cite{AK,Ch,ACK}.

The purpose of this paper is to address the constructibility of
theories of spin $1$ massless particles from a complementary
perspective. Instead of assuming a priori knowledge of Yang-Mills
Lagrangian with its Feynman diagrams, we here directly construct
candidates for the tree-level amplitudes via BCFW recursion
relations and then prove such a construction is consistent and the
resultant amplitudes are indeed the correct physical amplitudes,
given that some conditions on three-particle couplings are
satisfied.

Speaking specifically, Benincasa and Cachazo have discussed the
consistency conditions on four particle amplitudes constructed from
three particle ones for massless particles with general
spins~\cite{BC}. In particular, as for spin $1$ massless particles,
such consistency conditions require that the negative dimension
couplings should be absent and the dimensionless coupling constants
should be the structure constants of a Lie group. However, a
generalization to amplitudes with more external particles is still
lacking and the aim of this paper is to fill this gap. It turns out
that no further consistency conditions are needed for higher-point
amplitudes and BCFW construction automatically gives the correct
physical amplitudes with five or more particles. Therefore, along
with the result for four-particle test in~\cite{BC}, the present
paper provides the first purely S-matrix proof of constructibility
of theories for spin $1$ massless particles\footnote{The similar
result has also been obtained independently by Schuster and
Toro~\cite{ST}.}.

The paper is organized as following. After a brief review of
scattering amplitudes for massless particles and its construction
via BCFW recursion relations in Section \ref{pre}, we set out in
Section \ref{proof} to prove the consistency conditions on
scattering amplitudes of spin $1$ massless particles by induction.
Conclusion and discussions are presented in the end.
\section{Preliminaries}\label{pre}
\subsection{S-matrix of massless particles}
To set our notation and introduce convenient spinor language, we
review S-matrix first for general theories in four dimensional
Minkowski spacetime and then for theories with massless particles.

Probability for scattering process from asymptotic initial state to
final state is of particular physical interests, and it can be
calculated from physical inner-product of multi-particle states,
i.e.,
\begin{equation}
_{out}\langle
p_1,p_2,...,p_m|p'_1,p'_2,...,p'_{n-m}\rangle_{in}=\langle
p_1,p_2,...,p_m|S|p'_1,p'_2,...,p'_{n-m}\rangle,
\end{equation}
where $S=I+iT$ is an unitary operator. Then we can define the
scattering amplitude $M$ by
\begin{equation}
\langle
p_1,p_2,...,p_m|iT|p'_1,p'_2,...,p'_{n-m}\rangle=\delta^{4}(\sum_{i=1}^{m}p_i-\sum_{i=1}^{n-m}p'_i)M(\{p'_{1},...,p'_{n-m}\}\rightarrow\{p_{1},...,p_{m}\}).
\end{equation}

Instead of working with both ingoing and outgoing particles, we can
define scattering amplitude with only outgoing particles by using
$p_{m+1}=-p'_{1}$,$p_{m+2}=-p'_{2}$,...,$p_{n}=-p'_{n-m}$. As a
result, the probability of any process which involves $n$ particles
in total can be calculated by analytically continuing
$M_{n}(p_1,p_2...,p_n)$.

For any Poincare-invariant theory of massless particles in four
dimensional Minkowski spacetime, one-particle states, from which
multi-particle states are constructed, are irreducible massless
representations of Poincare group. An irreducible massless
representation is labeled by the on-shell momentum $p$ satisfying
$p^2=0$ and  helicity $h=\pm s$ where $s$ is the spin of the
particle. Furthermore, any on-shell momentum of massless particle
can be decomposed into
\begin{equation}
p^{\mu}=\lambda^a(\sigma^{\mu})_{a\dot{a}}\tilde{\lambda}^{\dot{a}}
\end{equation}
with $\sigma^{\mu}=(1,\sigma^{i})$. Of course this decomposition is
not unique, but only up to a little group transformation
$\lambda\rightarrow t\lambda,\tilde{\lambda}\rightarrow
t^{-1}\tilde{\lambda}$. In addition, it is noteworthy that for
complex momentum, which is essential for BCFW construction and
naturally defined by using complexified Lorentz group $SL(2,C)\times
SL(2,C)$, $\lambda$ and $\tilde{\lambda}$ are completely
independent. Note that any Lorentz invariants can be constructed
from basic invariants, i.e.,
\begin{equation}
\langle\lambda,\lambda'\rangle=\varepsilon^{ab}\lambda_{a}\lambda'_{b},[\tilde{\lambda},\tilde{\lambda}']=\varepsilon^{\dot{a}\dot{b}}\tilde{\lambda}_{\dot{a}}\tilde{\lambda}'_{\dot{b}}.
\end{equation}
An important example is the invariant from two on-shell momenta,
$p^{\mu}=\lambda\sigma^{\mu}\tilde{\lambda}$ and
$q^{\mu}=\lambda'\sigma^{\mu}\tilde{\lambda}'$, i.e.,
\begin{equation}
(p+q)^2=2p\cdot
q=\langle\lambda,\lambda'\rangle[\tilde{\lambda},\tilde{\lambda}'].
\end{equation}

All information of massless particles is encoded in pairs of spinors
and their helicities, from which an amplitude can be constructed.
This can be clearly seen from Feynman diagrams, where all
components, including propagators, vertices and polarization vectors
are functions of spinors and helicities. In particular, since an $n$
particle amplitude $M_n$ is Lorentz invariant, we conclude that it
is a function of $n$ helicities and basic Lorentz invariants
constructed from $n$ pairs of spinors, $\langle
i,j\rangle\equiv\langle \lambda_i,\lambda_j\rangle$ and
$[i,j]=[\tilde{\lambda}_i,\tilde{\lambda}_j]$ for $1\leq i,j\leq n$.
In addition, for tree-level amplitude, such a function is rational.
\subsection{BCFW construction}
The BCFW recursion relations can be schematically written as
\begin{eqnarray}\label{BCFW}
&&M_n^{(l,m)}(\{p_i,h_i,a_i\}|i=1,2,...,n)=\nonumber\\
&&\sum_{I,h,a}M_{|I|+1}(I(z_I),\{-P_I(z_I),-h,a\})\frac{1}{P^2_I}M_{|\bar{I}|+1}(\{P_I(z_I),h,a\},\bar{I}(z_I)).
\end{eqnarray}
Here some explanations are needed. We have picked two reference
particles $(l,m)$ with their momenta in sub-amplitudes deformed
as\footnote{The order of reference particles are relevant, i.e.,
$(l,m)$ and $(m,l)$ correspond to different deformations.},
\begin{equation}
\lambda^{(l)}(z)=\lambda^{(l)}+z\lambda^{(m)},\tilde{\lambda}^{(m)}(z)=\tilde{\lambda}^{(m)}-z\tilde{\lambda}^{(l)}.
\end{equation}
where for each subset $I\subseteq\{1,2,...,n\}$ with $l\in I$ and
$m\in \bar{I}$, the parameter $z$ is valued at the pole of the
amplitude, $z_I$, corresponding to sending the momentum of internal
legs $P_I(z_I)=\sum_{i\in I}p_i(z_I)=-\sum_{i\in\bar{I}}p_i(z_I)$ on
shell, with $p_i(z_I)=p_i$ undeformed for $i\neq l,m$. The summation
is over all divisions of external particles into $I$ and $\bar{I}$
with $l\in I$ and $m\in \bar{I}$, as well as the helicity $h$ and
color $a$ of internal leg. Every term in the summation is a product
of an $|I|+1$ particle sub-amplitude and an $|\bar{I}|+1$ particle
sub-amplitude where $|I|$ and $|\bar{I}|$ are numbers of external
particles in the subset $I$ and $\bar{I}$, and there is one on-shell
internal leg with $\{\mp P_I(z_I), \mp h, a\}$ for each of the
sub-amplitude, respectively. In addition, $P_I=\sum_{i\in
I}p_i=-\sum_{i\in\bar{I}}p_i$ is the off-shell momentum of the
propagator without deformation.

Details of the proof of BCFW recursion relations for amplitudes in
gauge theories and gravity, as well as more general theories can be
found in~\cite{BCFW,BBST,CS,BBC,H,AK,Ch,ACK}. Here we shall not
repeat the proof but only give some explanations. After the
deformation, the tree-level amplitude becomes a rational function of
$z$ and the key point for the proof is to show it vanishes as $z$
goes to infinity, i.e., $\lim_{z\rightarrow
\infty}M_n^{(l,m)}(z)=0$. Given this remarkable property, we have
$\oint_{C}M_n^{(l,m)}(z)/z=0$ when the contour $C$ encloses all
poles of the function, and the residue theorem implies
\begin{equation}
M_n^{(l,m)}\equiv M_n^{(l,m)}(z=0)=-\sum_{z_I}res_{z=z_I}
\frac{M_n^{l,m}(z)}{z},
\end{equation}
which essentially gives Eq.(\ref{BCFW}).

As pointed out before, in this paper we assume no a prior knowledge
on how to determine $M_n$ by the Yang-Mills Lagrangian with Feynman
diagrams and then check if it satisfies Eq.(\ref{BCFW}).
Alternatively, we consider Eq.(\ref{BCFW}) as our starting point to
\emph{define} a rational function $M_n^{(l,m)}$, which serves as a
candidate of the amplitude, and then prove the construction is
consistent and the rational function obtained is indeed the correct
amplitude $M_n$. An important remark is that the construction only
works for deformations on any two particles with helicities $(+,+)$,
$(+,-)$, and $(-,-)$, which we shall name as good deformations, but
not for the case $(-,+)$, which is called bad deformation. This has
been proved using Yang-Mills Lagrangian and its Feynman
diagrams~\cite{BCF,BCFW}. We shall also verify it below in our
purely S-matrix proof.
\section{Consistency conditions on tree-level amplitudes of spin $1$ massless
particles}\label{proof} In this section, we shall determine
consistency conditions for any tree-level amplitude to be
constructed from sub-amplitudes with fewer external particles using
BCFW construction. It turns out that non-trivial constraints only
come from consistency conditions on four particle amplitude
constructed from three particle ones, or the four-particle
test~\cite{BC}, which will be briefly summarized in \ref{initial}.
Once the test is passed, the correct physical $n$ particle amplitude
can be consistently constructed from lower amplitudes, which we
shall prove in \ref{proof1} and \ref{proof2}.

We prove this by induction. Suppose the consistency conditions are
satisfied for amplitudes with $4$,...,$n-1$ particles with $n\geq
5$. There are two different versions, a weak version and a strong
version, of these conditions. The strong version means that
$M^{(i,j)}_k=M^{(l,m)}_k$ for $k=4,...,n-1$ and any $1\leq
i,j,l,m\leq k$ as long as all deformations are good, and the
amplitude constructed this way has all correct factorization
channels, yielding the correct physical amplitudes. The weak version
only states that $M_k(1,2,...,k)$ can be constructed by lower
amplitudes using some deformations which give the same result, and
this is enough to ensure it to be the correct physical amplitude.

Here we only prove the weak version. Suppose we only have
$M_k^{(i,i-1)}=M_k^{(i,i+1)}$ and $M_k^{(i-1,i)}=M_k^{(i+1,i)}$ for
$1\leq i\leq k$, as long as all deformations involved are good, and
the amplitude constructed this way is the correct physical amplitude
for $k=4,...,n-1$. Then we shall prove that the weak version of
consistency conditions are also satisfied for $n$ particle
amplitudes, which is enough for our purpose.

In \ref{proof1} we shall prove $M_n^{(i,i-1)}=M_n^{(i,i+1)}$ and
$M_n^{(i-1,i)}=M_n^{(i+1,i)}$ for $1\leq i\leq n$, as long as all
deformations involved are good. The proof that these equalities have
guaranteed it to possess correct factorization channels, including
holomorphic and anti-holomorphic collinear limits, will be presented
in \ref{proof2}.
\subsection{Three particle amplitudes and four-particle test}\label{initial}
The starting point of~\cite{BC} is that three particle amplitudes
$M_3(\{p_i,h_i,a_i\}|i=1,2,3)$ for spin $s$ massless particles in
four dimensional Minkowski spacetime are completely determined by
Poincare symmetry, up to coupling constants. They are either
holomorphic or anti-holomorphic\footnote{The fact that these
amplitudes are either holomorphic or anti-holomorphic is simply due
to the physical condition that any three particle amplitude vanishes
for real momenta. For instance, when one takes the limit of real
momenta, if $h_1+h_2+h_3<0$, all anti-holomorphic coupling constants
$\kappa'_{a_1a_2a_3}$ must vanish to avoid a possible divergence.},
i.e.,
\begin{equation}\label{holomorphic}
M_3(\{p_i,h_i,a_i\}|i=1,2,3)=\kappa_{a_1a_2a_3}
\langle1,2\rangle^{d_3}\langle2,3\rangle^{d_1}\langle3,1\rangle^{d_2},
\end{equation}
for $h_1+h_2+h_3<0$, and
\begin{equation}\label{antiholomorphic}
M_3(\{p_i,h_i,a_i\}|i=1,2,3)={\kappa'}_{a_1a_2a_3}
[1,2]^{-d_3}[2,3]^{-d_1}[3,1]^{-d_2}
\end{equation}
for $h_1+h_2+h_3>0$. Here $d_1=h_1-h_2-h_3$, $d_2=h_2-h_3-h_1$,  and
$d_3=h_3-h_1-h_2$. In addition, $\kappa_{a_1a_2a_3}$ and
$\kappa'_{a_1a_2a_3}$ are coupling constants for particles with
colors $a_1$, $a_2$ and $a_3$, which can be separated into
dimensionless coupling constants $f_{a_1a_2a_3}$, and generically
dimensionful coupling constants $\kappa$ and $\kappa'$, which are
independent of color indices but can have helicity dependence. In
fact, a simple dimension analysis shows that both $\kappa$ and
$\kappa'$ have the dimension $1-|h_1+h_2+h_3|$ for the case of spin
$s$, which equals $1-3s$ for $+++$ and $---$ couplings and $1-s$ for
other cases. For $s=1$ we denote them as
$\kappa^{[-2]}$($\kappa'^{[-2]}$) and
$\kappa^{[0]}$($\kappa'^{[0]}$), respectively. A basic observation
on dimensionless coupling constants is that for odd $s$,
$f_{a_1a_2a_3}$ are antisymmetric with respect to any two subscripts
since in this case $d_1$,$d_2$ and $d_3$ are all odd.

Next one can build the four particle tree-amplitudes from three
particle ones by means of BCFW recursion relations. However, as
shown in~\cite{BC}, one needs consistency condition on the
amplitude, i.e., four particle test: different constructions by
deforming particles $(1,2)$ and $(1,4)$ must give the same result
$M_4^{(1,2)}=M_4^{(1,4)}$. This simple condition imposes severe
constraints on non-trivial theories with non-zero coupling
constants, which can be summarized as\footnote{We do not repeat the
proof here, and details can be found in~\cite{BC}.}: (1)
$\kappa^{[-2]}=\kappa'^{[-2]}=0$, i.e., there is no $+++$ or $---$
coupling, (2) the dimensionless coupling constants must conform to
Jacobi condition, i.e.,
$\sum_{e}(f_{ade}f_{ebc}+f_{ace}f_{edb}+f_{abe}f_{ecd})=0$.

As will become clear, the first constraint is crucial for our proof
in \ref{proof1}. In~\cite{BC}, this has also been shown to come from
the condition for constructibility by analysis of Lagrangian and
Feynman diagrams. From the Lagrangian point of view, this excludes
higher derivative terms like $(F^2)^2$. Here we want to stress that
this constraint is part of the consistency conditions on amplitudes
constructed by BCFW recursion relations, which holds without the
assumption of Lagrangian and Feynman diagrams.

Let us focus on the second constraint, which, together with the fact
that each $f_{abc}$ is totally antisymmetric, implies that $f_{abc}$
constitute the structure constants of a Lie algebra. We shall assume
in the following that the Lie algebra is
$su(\mathfrak{n})$\footnote{Note that this $\mathfrak{n}$ has
nothing to do with $n$, the number of external particles in
amplitudes.}, which is our main interest. Suppose $T^a$ to be the
generators of $su(\mathfrak{n})$, which satisfy $[T^a,T^b]=f_{abc}
T^c$. Since we have assumed consistency conditions on
$M_k(1,2,...,k)$ for $k=3,...,n-1$, which guarantee them to be the
correct physical amplitudes, it is well known, at least for
$su(\mathfrak{n})$, that we can do the color decomposition for any
tree amplitudes,
\begin{equation}\label{colordk}
M_k(\{p_i,h_i,a_i\}|i=1,...,k)=\sum_{\sigma\in
S_k/C_k}Tr(T^{a_{\sigma(1)}}...T^{a_{\sigma(k)}})M^P_k(\sigma(1^{h_1}),...,\sigma(k^{h_k})).
\end{equation}
for $k=3,...,n-1$. Here $S_k$ is the permutation group and $C_k$ is
the corresponding cyclic subgroup. In addition,
$M^P_k(1^{h_1},...,k^{h_k})$, with $i^{h_i}$ referring to
$\{p_i,h_i\}$, are called the color-ordered amplitudes, or partial
amplitudes. We want to show that the same decomposition can also be
done for tree-level $n$ particle amplitudes constructed by recursion
relations for $n\geq 4$, using any good deformation, i.e.,
\begin{equation}\label{colordn}
M_n^{(l,m)}(\{p_i,h_i,a_i\}|i=1,...,n)=\sum_{\sigma\in
S_n/C_n}Tr(T^{a_{\sigma(1)}}...T^{a_{\sigma(n)}}){M^P}^{(l,m)}_n(\sigma(1^{h_1}),...,\sigma(n^{h_n})).
\end{equation}
The key point to justify Eq.(\ref{colordn}) is the identity for
$su(\mathfrak{n})$
\begin{equation}\label{su_n}
\sum_{a_I}Tr(T^{a_{1}}...T^{a_{i}}T^{a_{I}})Tr(T^{a_{I}}T^{a_{i+1}}...T^{a_{n}})=Tr(T^{a_{1}}...T^{a_{n}}).
\end{equation}
Since any lower amplitudes, from which the L.H.S. of
Eq.(\ref{colordn}) is constructed, can be decomposed as in
Eq.(\ref{colordk}), then by plugging Eq.(\ref{colordk}) into
Eq.(\ref{BCFW}) and using Eq.(\ref{su_n}) to carry out the summation
over $a_I$, we arrive at the R.H.S. of Eq.(\ref{colordn}), where
${M^P}^{(l,m)}_n$ is constructed from lower partial amplitudes and
eventually $M^P_3$. In addition, combining the $k=3$ case of
Eq.(\ref{colordk}) with Eq.(\ref{holomorphic}) or
Eq.(\ref{antiholomorphic}) shows that $M^P_3$ is just the R.H.S. of
Eq.(\ref{holomorphic}) with coupling constants $\kappa_{a_1a_2a_3}$
replaced by $\kappa$,or the R.H.S. of Eq.(\ref{antiholomorphic})
with $\kappa'_{a_1a_2a_3}$ replaced by $\kappa'$.

These partial amplitudes actually contain all the kinematic
information, and will be the major objects for study in the
following. So we henceforth omit the superscript $P$. Such partial
amplitudes are much simpler than the full ones due to several
reasons. First of all, they are cyclic-symmetric for its $n$
external legs, i.e.,
\begin{equation}
M_n(p_1,p_2,...,p_n)=M_n(p_{\sigma(1)},p_{\sigma(2)},...,p_{\sigma(n)}).
\end{equation}
for any $\sigma\in C_n$.

More importantly, they only receive contributions from diagrams with
a certain cyclic ordering of the external legs. An important
observation is that all poles of these partial amplitudes merely
come from those channels with adjacent momenta, like
$s_{i,...,j}=(p_i+p_{i+1}...+p_{j-1}+p_j)^2$. This thus vastly
reduced the number of terms that can appear in their BCFW
construction. For example, if we want to use the recursion relations
by deforming two adjacent particles of an $n$ particle partial
amplitude, say $(1,n)$, all divisions of the set $\{1,2,...,n\}$ we
need to consider are only those of the form $I=\{1,2,...i\}$ and
$\bar{I}=\{i+1,...,n\}$, with $n-3$ terms, i.e., $2\leq i\leq n-2$,
instead of any subsets $I$ with $1\in I$ and $n\in \bar{I}$, which
is the case for full amplitudes. The result can be schematically
summarized as
\begin{eqnarray}
&&M_n^{(n,1)}(p_i,h_i|i=1,2,...,n)=\nonumber\\
&&\sum_{i=2}^{n-2}\sum_{h=\pm}M_{i+1}(1(z_i),...,i,\{-P_i(z_i),-h\})\frac{1}{P_i^2}M_{n-i+1}(\{P_i(z_i),h\},i+1,...,n(z_i)).\nonumber\\
\end{eqnarray}
\subsection{A note on notations}
Before proceeding to prove our consistency conditions, we would like
to pause a moment to fix our notations, which will make our formula
compact.

\begin{figure}[htbp]
\includegraphics[width=\textwidth]{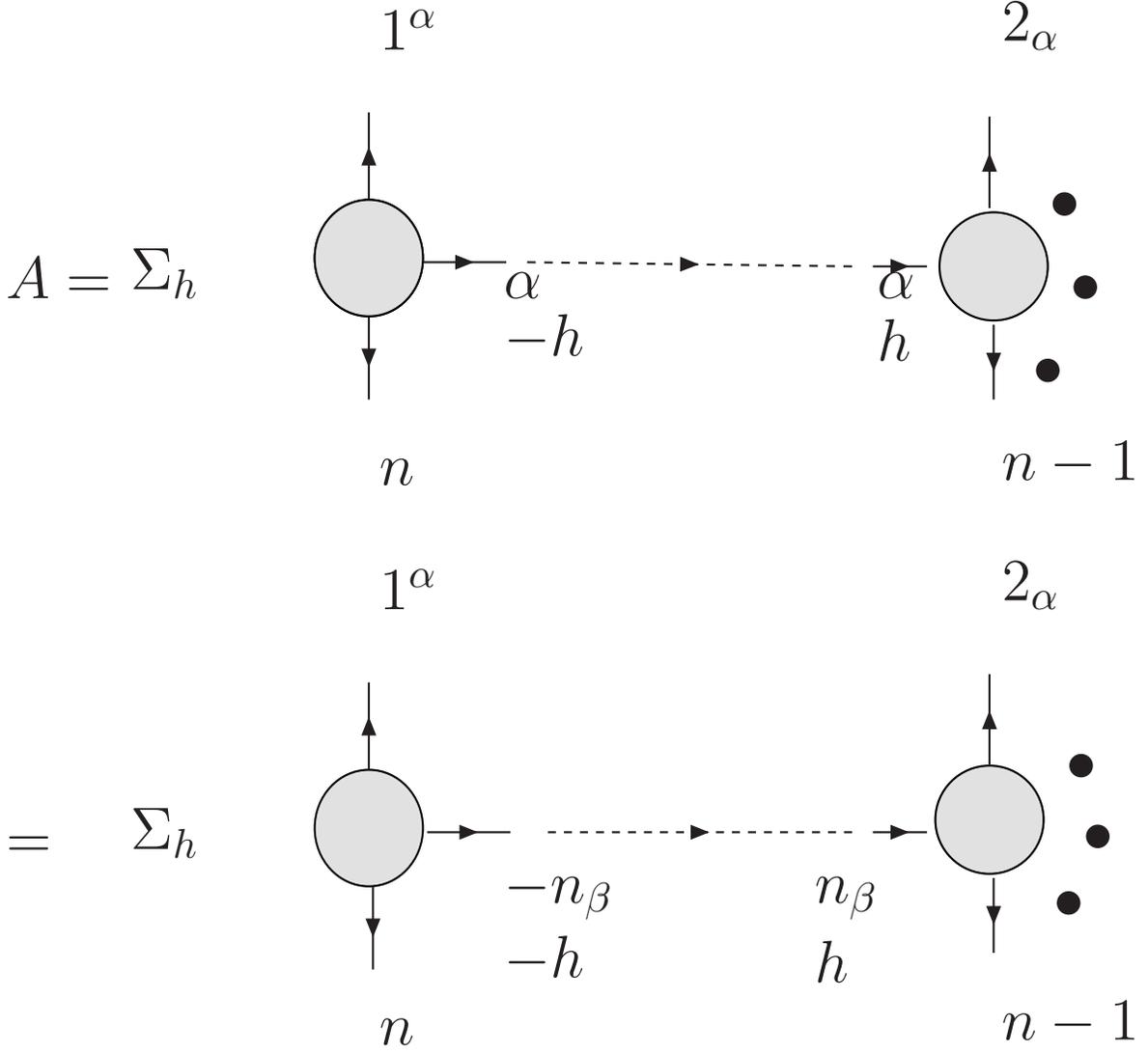}
\caption{Terms in $M_n^{(1,2)}$ with particle 1 in three amplitudes,
where dots denote other external particles and dashed lines are
off-shell propagators, $\alpha$ and $\beta$ are short for
$\alpha_{n-1}$ and $\beta_{n-1}$. In the second line, we use
$n_\beta=n\oplus1^\alpha$ for internal legs.} \label{figA}
\end{figure}

\begin{figure}[htbp]
\includegraphics[width=\textwidth]{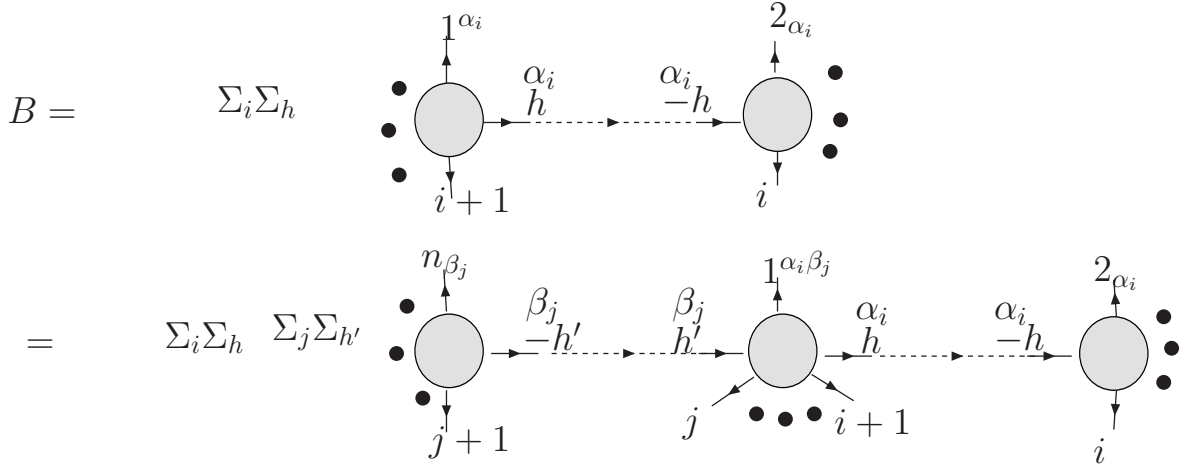}
\caption{Other terms in $M_n^{(1,2)}$ where dots denote other
external particles and dashed lines are off-shell propagators. In
the second line we further factorize the left amplitude by deforming
the pair $(1,n)$.} \label{figB}
\end{figure}

\begin{figure}[htbp]
\includegraphics[width=\textwidth]{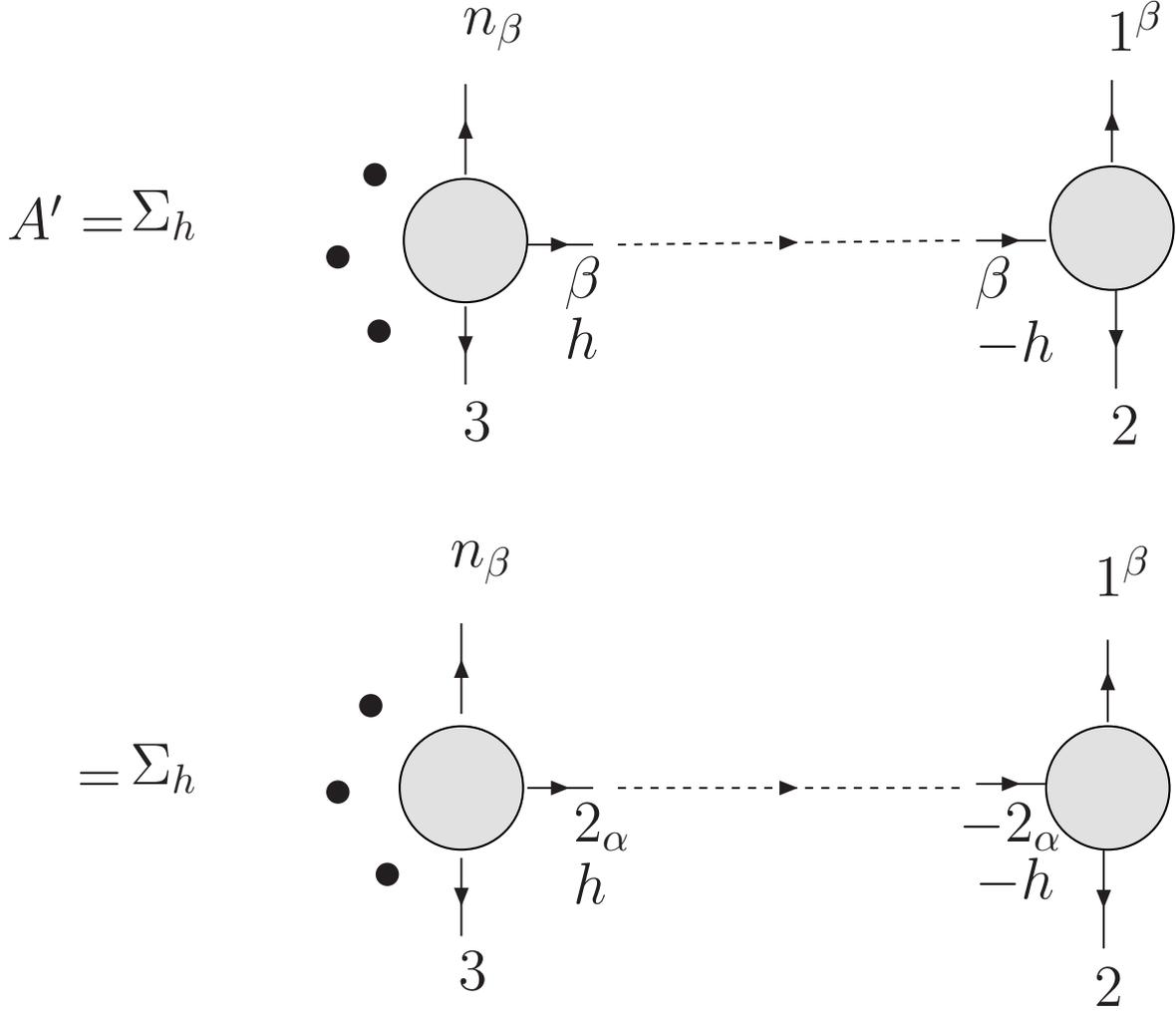}
\caption{Terms in $M_n^{(1,n)}$ with particle 1 in three amplitudes,
where dots denote other external particles and dashed lines are
off-shell propagators, $\alpha$ and $\beta$ are short for
$\alpha_{n-1}$ and $\beta_{n-1}$. In the second line we use
$2_\alpha=1^\beta\oplus2$ for internal legs, then the left amplitude
is the same as that in A.} \label{figA'}
\end{figure}

\begin{figure}[htbp]
\includegraphics[width=\textwidth]{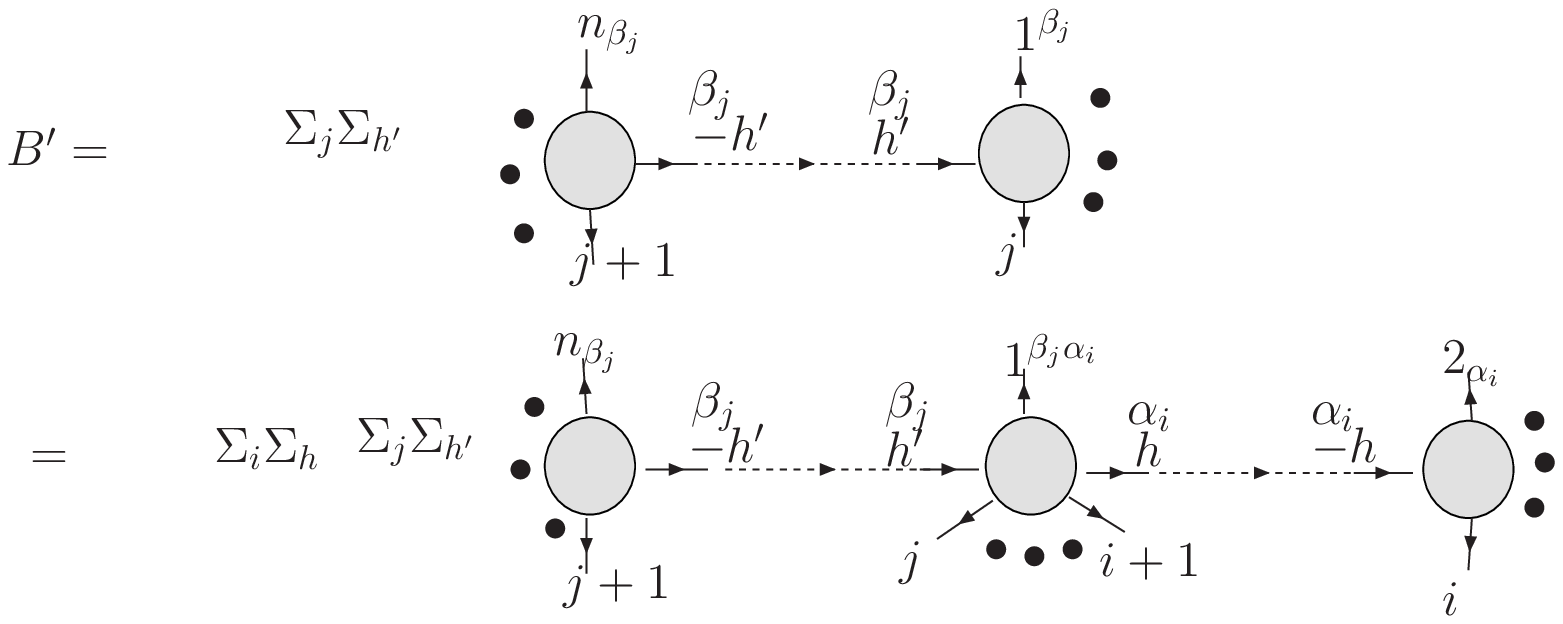}
\caption{Other terms in $M_n^{(1,n)}$ where dots denote other
external particles and dashed lines are off-shell propagators. In
the second line we further factorize the right amplitude by
deforming the pair $(1,n)$, then the second line is the same as that
in B.} \label{figB'}
\end{figure}

Following~\cite{BC}, we denote a pair of spinors
$\{\lambda^{(i)},\tilde{\lambda}^{(i)}\}$ corresponding to an
on-shell momentum $p_i$ by $i$ for $i=1,...,n$. Now in a deformation
on $(i,j)$, we use a Greek letter as superscript of $i$ to denote
the left-handed spinor of $i$ being shifted, while the same letter
as subscript of $j$ is used to denote the right-handed one of $j$
being shifted. Deformations on different pairs of particles will be
represented by different Greek letters. For example, as illustrated
in Figure \ref{figA}, \ref{figB}, \ref{figA'} and \ref{figB'}, the
deformation on $(1,2)$ results in $1^\alpha$ and $2_\alpha$, while
the one on $(1,n)$ yields $1^\beta$ and $n_\beta$.

Furthermore, in both sub-amplitudes of a factorization, momenta are
understood to be deformed with the parameter $z$ at the pole of the
original amplitude, which keeps momenta of internal legs in this
factorization on-shell, as required by the recursion relations.
Therefore different factorizations from deforming the same pair of
particles have different parameters of deformations, and to label
momenta in these factorizations, we need to add subscripts
representing different factorizations to the same Greek letter,
which are shown in Figure \ref{figA},\ref{figB}, \ref{figA'} and
\ref{figB'}, where $\alpha$ and $\beta$ are short for $\alpha_{n-1}$
and $\beta_{n-1}$.

For those on-shell internal legs, we use $i^\alpha\oplus...\oplus j$
to represent a pair of spinors whose momentum is given by
$P=p_{i^\alpha}+...+p_j$(up to little group transformation), while
the pair of spinors representing $P=-(p_{i^\alpha}+...+p_j)$ is
denoted by $-(i^\alpha\oplus...\oplus j)$. Momentum conservation
ensures that we can use either $i^\alpha\oplus...\oplus j$ or
$-(k_\alpha\oplus...\oplus l)$ for an internal leg from deformation
on $(i,k)$, where $i,...,j$ are all other particles in the same
sub-amplitude, and $k,...,l$ are all particles except the internal
leg in the sub-amplitude on the other side of the propagator. We
also explicitly use $\pm h$ to represent opposite helicities of
internal legs on two sides of the propagator. Finally,
$(p_i+...+p_j)^2$ is denoted by $|i\oplus...\oplus j|^2$ in the
propagator.

\subsection{Proof of $M_n^{(i,i-1)}=M_n^{(i,i+1)}$ and $M_n^{(i-1,i)}=M_n^{(i+1,i)}$}\label{proof1}
The first step is to prove
$M_n^{(1^\alpha,2_\alpha)}=M_n^{(1^\beta,n_\beta)}$. According to
the BCFW construction, the $n$ particle partial amplitude with
particle $1$ and $2$ deformed can be constructed as,
\begin{eqnarray}\label{Mn12}
&&M_n^{(1^\alpha,2_\alpha)}=\sum_{h=\pm}M_3(n,1^{\alpha_{n-1}},\{-(n\oplus1^{\alpha_{n-1}}),-h\})\frac{1}{|n\oplus1|^2}\nonumber\\
&&\times M_{n-1}(\{n\oplus1^{\alpha_{n-1}},h\},2_{\alpha_{n-1}},3,...,n-1)\nonumber\\
&&+\sum^{n-2}_{i=3}\sum_{h=\pm}[M_{n-i+2}(i+1,...,n,1^{\alpha_{i}},\{2_{\alpha_{i}}\oplus...\oplus i,h\})\frac{1}{|2\oplus...\oplus i|^2}\nonumber\\
&&\times M_{i}(\{-(2_{\alpha_{i}}\oplus...\oplus
i),-h\},2_{\alpha_{i}},...,i)].
\end{eqnarray}
Here we have divided the sum over different ways of factorizations
into two parts, the former of which corresponding to $i=n-1$ is
denoted by $A$ while the latter, the sum over $i$ from $3$ to $n-2$,
is denoted as $B$. These two terms are shown in the first lines of
figure \ref{figA} and figure \ref{figB}, respectively. As mentioned
before, we add $i$ as subscripts to $\alpha$ since parameters
$z(\alpha_i)$ of deformations are different for different
factorizations.

Similarly, the amplitude with particle $1$ and $n$ deformed can also
be constructed, i.e.,

\begin{eqnarray}\label{Mn1n}
&&M_n^{(1^\beta,n_\beta)}=\sum_{h=\pm}M_{n-1}(3,...,n-1,n_{\beta_{n-1}},\{1^{\beta_{n-1}}\oplus2,h\})\frac{1}{|1\oplus2|^2}\nonumber\\
&&\times M_{3}(\{-(1^{\beta_{n-1}}\oplus2),-h\},1^{\beta_{n-1}},2)\nonumber\\
&&+\sum^{n-2}_{j=3}\sum_{h'=\pm}[M_{n-j+1}(j+1,...,n_{\beta_j},\{-((j+1)\oplus...\oplus n_{\beta_j}) ,h'\})\frac{1}{|1\oplus...\oplus j|^2}\nonumber\\
&&\times M_{j+1}(\{(j+1)\oplus...\oplus
n_{\beta_j},-h'\},1^{\beta_j},2,...,j)],
\end{eqnarray}
where the term in the first two lines with $j=2$ is denoted as
$A'$ and the rest, the sum over $j$ from $3$ to $n-2$, is denoted by
$B'$. These are shown in first lines of figure \ref{figA'} and
figure \ref{figB'}, respectively.

To proceed, first we notice that by deformation on
$(1^{\alpha_i},n)$, as illustrated in the second line of figure
\ref{figB}, $B$ can be factorized further as
\begin{eqnarray}\label{B}
&&B=\sum^{n-2}_{i=3}\sum^{n-2}_{j=i}\sum_{h=\pm}\sum_{h'=\pm}[M_{n-j+1}(j+1,...,n_{\beta_j},\{-((j+1)\oplus...\oplus n_{\beta_j}),-h'\})\nonumber\\
&&\times\frac{1}{|(j+1)\oplus...\oplus n|^2}M_{j-i+3}(\{(j+1)\oplus...\oplus n_{\beta_j},h'\},{1^{\alpha_i}}^{\beta_j},\{2_{\alpha_i}\oplus...\oplus i,h\},i+1,...,j)\nonumber\\
&&\times\frac{1}{|2\oplus...\oplus
i|^2}M_{i}(\{-(2_{\alpha_i}\oplus...\oplus
i),-h\},2_{\alpha_i},...,i)].
\end{eqnarray}
Here in the summation over $j$ from $i$ to $n-2$, when $j=i$, the
sub-amplitude $M_{j-i+3}$ is just a three particle amplitude and
external legs $\{i+1,...,j=i\}$ in it are understood to be an empty
set in this case.

An important observation is that here we can use $\beta_j$ with
$j=i,...,n-2$ for these deformations on $(1^{\alpha_i},n)$ just as
those deformations on $(1,n)$, which can be justified as following.
Suppose we denote these deformations with super(sub)scripts $\mu_j$.
Note that $\tilde{\lambda}^{(1^{\alpha_i})}=\tilde{\lambda}^{(1)}$
and $\lambda^{(n_{\mu_j})}=\lambda^{(n)}$. Then
$\tilde{\lambda}^{(n_{\mu_j})}=\tilde{\lambda}^{(n)}-z(\mu_j)\tilde{\lambda}^{(1)}$
where the parameter $z(\mu_j)$ for a factorization $j$ is determined
by the on-shell condition of $(j+1)\oplus...\oplus n_{\mu_j}$ which
gives exactly the same equation for $z(\mu_j)$ as that of
$z(\beta_j)$. So we have $z(\mu_j)=z(\beta_j)$, which further
implies
$\tilde{\lambda}^{(n_{\mu_j})}=\tilde{\lambda}^{(n_{\beta_j})}$ and
$\lambda^{({1^{\alpha_i}}^{\mu_j})}=\lambda^{(1^{\alpha_i})}+z(\beta_j)\lambda^{(n)}=\lambda^{({1^{\alpha_i}}^{\beta_j})}$,
where ${1^{\alpha_i}}^{\beta_j}$ is understood as the composition of
two deformations on particle $1$, with the left one done first.

Similarly, as shown in the second line of figure \ref{figB'}, $B'$
can be factorized by deforming $(1^{\beta_j},2)$,
\begin{eqnarray}\label{B'}
&&B'=\sum^{n-2}_{j=3}\sum^{j}_{i=3}\sum_{h=\pm}\sum_{h'=\pm}[M_{n-j+1}(j+1,...,n_{\beta_j},\{-((j+1)\oplus...\oplus n_{\beta_j}),-h'\})\nonumber\\
&&\times\frac{1}{|(j+1)\oplus...\oplus n|^2}M_{j-i+3}(\{(j+1)\oplus...\oplus n_{\beta_j},h'\},{1^{\beta_j}}^{\alpha_i},\{2_{\alpha_i}\oplus...\oplus i,h\},i+1,...,j)\nonumber\\
&&\times\frac{1}{|2\oplus...\oplus
i|^2}M_{i}(\{-(2_{\alpha_i}\oplus...\oplus
i),-h\},2_{\alpha_i},...,i)].
\end{eqnarray}
Here we justify the use of $\alpha_i$ for the same reason as before.
In ${1^{\beta_j}}^{\alpha_i}$, the order of actions of two
deformations on particle $1$ are reversed from that in
${1^{\alpha_i}}^{\beta_j}$. In both cases, the right-handed spinor
$\tilde{\lambda}^{(1)}$ remains unchanged, while
$\lambda^{({1^{\beta_j}}^{\alpha_i})}=\lambda^{(1)}+z(\beta_j)\lambda^{(n)}+z(\alpha_i)\lambda^{(2)}=\lambda^{({1^{\alpha_i}}^{\beta_j})}$.
So we have ${1^{\beta_j}}^{\alpha_i}={1^{\alpha_i}}^{\beta_j}$. In
addition, note that the summation over $i$ and $j$ in both
Eq.(\ref{B}) and Eq.(\ref{B'}) is actually the summation over all
$i,j$ with $3\leq i\leq j\leq n-2$. Therefore, as shown in figures
\ref{figB} and \ref{figB'}, we conclude that $B=B'$.

Now we proceed to prove $A=A'$. Since in $A$ and $A'$ we only
encounter $\alpha_{n-1}$ and $\beta_{n-1}$, we henceforth omit the
subscript $n-1$ and use $\alpha$ and $\beta$, just as we did in
figures\ref{figA} and \ref{figA'}. By
Eq.(\ref{alpha}),Eq.(\ref{beta}),Eq.(\ref{noplus1alpha}), and
Eq.(\ref{1betaoplus2}) in the Appendix, we obtain
$n\oplus1^\alpha=n_\beta$ and $1^\beta\oplus 2=2_\alpha$. Taking
into account $|i\oplus j|^2=\langle i,j\rangle[i,j]$, we have the
following simplified expressions for $A$ and $A'$,
\begin{eqnarray}\label{AA'}
A&=&\sum_{h<h_n+h_1}M_3(n^{h_n},(1^\alpha)^{h_1},(-n_\beta)^{-h})\frac{1}{\langle n,1\rangle[n,1]}M_{n-1}(n_\beta^h,2_\alpha^{h_2},3^{h_3},...,(n-1)^{h_{n-1}}),\nonumber\\
A'&=&\sum_{h<h_1+h_2}M_3((1^\beta)^{h_1},2^{h_2},(-2_\alpha)^{-h}\})\frac{1}{\langle1,2\rangle[1,2]}M_{n-1}(n_\beta^{h_n},2_\alpha^{h},3^{h_3},...,(n-1)^{h_{n-1}}),
\end{eqnarray}
which are shown in second lines of Figure \ref{figA} and Figure
\ref{figA'}, respectively. Here we have recovered helicities for all
legs and taken into account the fact that terms with $h_n+h_1-h<0$
vanish in $A$ and those with $h_1+h_2-h<0$ vanish in
$A'$.\footnote{This is because if, say in $A'$, $h_1+h_2-h<0$, then
the three particle amplitude must be holomorphic. However, both
$\lambda^{1^{\beta}}$ and $\lambda^{2_{\alpha}}$ are proportional to
$\lambda^{2}$, thus the amplitude possesses a factor
$\langle2,2\rangle^{-h_1-h_2+h}$, which vanishes.}

To proceed, first we notice from Eq.(\ref{AA'}) that two $n-1$
particle amplitudes in $A$ and $A'$ are almost the same, except that
the first two helicities, $(h,h_2)$ in $A$ and $(h_n,h)$ in $A'$ may
not be the same. So we shall only keep these two variables in
$M_{n-1}$ for both $A$ and $A'$ below. For three particle
amplitudes, one needs Eq.(\ref{holomorphic}) and
Eq.(\ref{antiholomorphic}), where for illustration we shall keep
$\kappa^{[-2]}$ and $\kappa'^{[-2]}$ , although the four-particle
test requires that both of them should vanish.

Now we are ready to discuss all possible helicity arrangements for
particle $1$, $2$ and $n$. If $h_1=h_n=-$, $A$ vanishes since there
is no term with $h_n+h_1-h>0$, so does $A'$ for $h_1=h_2=-$.
Therefore, given $B=B'$, we conclude that in the case
$(h_n,h_1,h_2)=(-,-,-)$, such two good deformations on $(1,2)$ and
$(1,n)$ produce the same result.

If $h_1=-$ and $h_n=+$, then we must have $h=-$ in $A$. So we can
obtain $A$ by Eq.(\ref{antiholomorphic}) as
\begin{equation}
A=\kappa'^{{[0]}}\frac{\langle1,2\rangle^3}{\langle
n,2\rangle^3}\langle n,1\rangle^{-1}M_{n-1}(-,h_2).
\end{equation}
Likewise, if $h_1=-$ and $h_2=+$, then we have $h=-$ in $A'$, which
gives
\begin{equation}
A'=\kappa'^{{[0]}}\frac{\langle n,1\rangle^3}{\langle
n,2\rangle^3}\langle 1,2\rangle^{-1}M_{n-1}(h_n,-).
\end{equation}

Given $B=B'$, for $(h_n,h_1,h_2)=(-,-,+)$, the bad deformation on
$(1,2)$ with $(-,+)$ gives vanishing $A$, which is different from
generically non-vanishing $A'$ given by good deformation on $(1,n)$
with $(-,-)$. Similarly, for $(h_n,h_1,h_2)=(+,-,-)$, the bad
deformation on $(1,n)$ with $(-,+)$ gives a different answer from
that obtained by the good deformation on $(1,2)$ with $(-,-)$.
Finally, for $(h_n,h_1,h_2)=(+,-,+)$, the two deformations on
$(1,2)$ and $(1,n)$ are both bad ones with helicities $(-,+)$, which
give different answers from each other generically. Therefore, as
promised, using purely S-matrix arguments, we have also derived
that the bad deformation with helicities $(-,+)$ can not be used in
BCFW construction.

For $h_1=+$, the deformations on $(1,2)$ and $(1,n)$ are always
good. After some algebraic calculations, the corresponding result
can be obtained as
\begin{equation}
A=\kappa'^{{[0]}}\frac{\langle n,2\rangle}{\langle1,2\rangle}\langle
n,1\rangle^{-1}M_{n-1}(h_n,h_2)+\delta_{h_n,+}\kappa'^{{[-2]}}\frac{\langle1,2\rangle}{\langle
n,2\rangle}[n,1]^{2}\langle n,1\rangle^{-1}M_{n-1}(-,h_2),
\end{equation}
and
\begin{equation}
A'=\kappa'^{{[0]}}\frac{\langle n,2\rangle}{\langle
n,1\rangle}\langle
1,2\rangle^{-1}M_{n-1}(h_n,h_2)+\delta_{h_2,+}\kappa'^{{[-2]}}\frac{\langle
n,1\rangle}{\langle
n,2\rangle}[1,2]^{2}\langle1,2\rangle^{-1}M_{n-1}(h_n,-),
\end{equation}
where we can immediately recognize that the terms with
$\kappa'^{{[0]}}$ are equal to each other while those with
$\kappa'^{{[-2]}}$ are generally not. Therefore, given $B=B'$, if
$(h_n,h_1,h_2)=(-,+,-)$, there are only first terms in both $A$ and
$A'$, so we get the same result for such two good deformations. For
$(h_n,h_1,h_2)=(-,+,+)$, $(+,+,-)$, or $(+,+,+)$, the two good
deformations yield the same answer if and only if
$\kappa'^{{[-2]}}=0$.

Therefore, given $\kappa'^{{[-2]}}=0$, which has been guaranteed by
the four-particle test, we conclude that, as long as no bad
deformation is involved,
\begin{equation}\label{equality}
M_n^{(1^\alpha,2_\alpha)}=M_n^{(1^\beta,n_\beta)}.
\end{equation}

Now consider
$M_n^{(2^{\alpha'},1_{\alpha'})}=M_n^{(n^{\beta'},1_{\beta'})}$. If
all deformations are good, the proof of this equality goes exactly
the same way as the proof of Eq.(\ref{equality}), only with all
helicities flipped and $\lambda\leftrightarrow\tilde{\lambda}$.

Finally, since any
partial amplitude is cyclic symmetric, our proof can be applied to
any leg $i$ of an $n$ particle partial amplitude, $1\leq i\leq n$,
as long as any deformation involved is good.

To summarize, we have proved that
$M_n^{(i^\mu,(i-1)_\mu)}=M_n^{(i^\nu,(i+1)_\nu)}$ and
$M_n^{((i-1)^{\mu'},i_{\mu'})}=M_n^{((i+1)^{\nu'},i_{\nu'})}$ hold
if and only if the deformation involved is good one.
\subsection{Proof of the correct factorizations of amplitudes}\label{proof2}
The final step is to show that the amplitude constructed by
deforming adjacent particles is the correct physical amplitude. As
discussed before, it is sufficient to check if the amplitude has all
the correct factorization channels, which for partial amplitudes are
only made up of adjacent momenta.

The statement that the amplitude has correct factorizations means if
we send the momentum of a channel on-shell, the amplitude should
contain a singular term which is the product of two sub-amplitudes
with the propagator of this channel, plus other non-singular terms.
We have supposed this is true for $M_k$ with $3\leq k\leq n-1$, and
now we prove that the $n$ particle partial amplitude constructed by
recursion relations also have correct factorizations for any channel
being sent on-shell.

Suppose we obtain the amplitude by deforming $(1,n)$, which is given
by Eq.(\ref{Mn1n}). Then any propagator appearing in Eq.(\ref{Mn1n})
comes from the channel in the form $s_{j,...,k}$ with $1\leq j<k\leq
n-2$ or $3\leq j<k \leq n$. We now want to check that, if one sends
the momentum of such a channel on-shell, i.e.,
$s_{j,...,k}\rightarrow0$, the amplitude really becomes a product of
two sub-amplitudes, with the singular propagator of this channel,
plus other non-singular terms.

First, we know that sub-amplitudes have the correct factorization
when $s_{j,...,k}\rightarrow 0$, i.e.,
\begin{eqnarray}\label{lower1}
&&M_{i+1}(1^\beta,...,j,...,k,...,i,-(1^\beta\oplus...\oplus i))=M_{k-j+2}(j,...,k,-(j\oplus...\oplus k))\nonumber\\
&&\times\frac{1}{|j\oplus...\oplus k|^2}M_{i-k+j+1}(1^\beta,...,j\oplus...\oplus k,...,i,-(1^\beta\oplus...\oplus i))\nonumber\\
&&+\textit{non-singular terms}
\end{eqnarray}
for $1\leq j<k\leq i\leq n-2$, and
\begin{eqnarray}\label{lower2}
&&M_{n-i+1}(i+1,...,j,...,k,...,n_\beta,-((i+1)\oplus...\oplus n_\beta))=M_{k-j+2}(j,...,k,-(j\oplus...\oplus k))\nonumber\\
&&\times\frac{1}{|j\oplus...\oplus k|^2}M_{n-i-k+j+1}(i+1,...,j\oplus...\oplus k,...,n_\beta,(i+1)\oplus...\oplus n_\beta)\nonumber\\
&&+\textit{non-singular terms}
\end{eqnarray}
for $2\leq i<j<k\leq n$.

Then by Eq.(\ref{Mn1n}), Eq.(\ref{lower1}), and Eq.(\ref{lower2}),
we obtain
\begin{eqnarray}
&&M_{n}^{(1^\beta,n_\beta)}=[\sum_{i=k}^{n-2}M_{n-i+1}(i+1,...,n_\beta,-((i+1)\oplus...\oplus n_\beta))\frac{1}{|(i+1)\oplus...\oplus n|^2}\nonumber\\
&&\times M_{i-k+j+1}(1^\beta,...,j\oplus...\oplus k,...,i,(i+1)\oplus...\oplus n_\beta)\nonumber\\
&&+\sum_{i=2}^{j-1}M_{i+1}(1^\beta,...,i,-(1^\beta\oplus...\oplus i))\frac{1}{|1\oplus...\oplus i|^2}\nonumber\\
&&\times M_{n-i-k+j+1}(i+1,...,j\oplus...\oplus k,...,n_\beta,1^\beta\oplus...\oplus i)]\nonumber\\
&&\times\frac{1}{|j\oplus...\oplus
k|^2}M_{k-j+2}(j,...,k,-(j\oplus...\oplus k))+\textit{non-singular
terms}
\end{eqnarray}
for $1\leq j<k\leq i\leq n-2$, or  $2\leq i<j<k\leq n$. Note that
all the terms in $[\ ]$, i.e., those from the first to the fourth
line, are the factorizations by deforming $(1,n)$ of an $n-k+j$
particle amplitude, which are obtained by replacing all the legs
from $j$ to $k$ with a single on-shell leg $j\oplus...\oplus k$,
thus we have
\begin{eqnarray}
&&M_{n}^{(1^\beta,n_\beta)}=M_{n-j+k}(1,...,j-1,j\oplus...\oplus k,k+1,...,n)\nonumber\\
&&\times\frac{1}{|j\oplus...\oplus
k|^2}M_{k-j+2}(j,...,k,-(j\oplus...\oplus k))+\textit{non-singular
terms},
\end{eqnarray}
which means the $n$ particle partial amplitude also has the correct
factorization when $s_{j,...,k}\rightarrow 0$ for $1\leq j<k\leq
i\leq n-2$, or $2\leq i<j<k\leq n$. Notice that
$s_{j,...,k}=s_{k+1,...,n,1,...,j-1}$, so the correct factorization
channels include all possible channels of the partial amplitude
except $s_{n,1}$.

However, an important thing we need to check is the inclusion of
both holomorphic and anti-holomorphic collinear limits. Since
$p_{i,i+1}^2=\langle i,i+1\rangle[i,i+1]$, we need to take care of
two separate cases\footnote{An example for illustration is a five
particle amplitude with certain helicity configuration, i.e., $
M_5\propto\frac{\langle1,2\rangle[3,4]}{[1,2]\langle3,4\rangle}$,
where it is easy to see that for real collinear limits, i.e., as
both $\langle1,2\rangle$ and $[1,2]$ go to zero, this function is
not singular. In fact there is no real collinear limit or
factorization that can detect this. However, an anti-holomorphic
factorization limit, i.e., $[1,2]\rightarrow 0$ while
$\langle1,2\rangle\neq 0$, detects it and the inclusion of this
anti-holomorphic collinear limit is needed for the function to be
the correct physical amplitude.}, i.e., the holomorphic pole,
$\langle i,i+1\rangle\rightarrow0$ while $[i,i+1]\neq 0$, and
anti-holomorphic pole,$[i,i+1]\rightarrow0$ while $\langle
i,i+1\rangle\neq0$.  From Eq.(\ref{Mn1n}), we can see that
$M_n^{(1^\beta,n_\beta)}$ has the correct factorizations at the
anti-holomorphic pole from the channel $s_{1,2}$, the holomorphic
pole from channel $s_{n-1,n}$, and at both holomorphic and
anti-holomorphic poles from all other channels except $s_{n,1}$.

In other words, we have not shown that $M_n^{(1^\beta,n_\beta)}$
given by Eq.(\ref{Mn1n}) has the correct factorizations at the
anti-holomorphic pole from $s_{n-1,n}$, the holomorphic pole from
$s_{1,2}$, and both poles from the channel $s_{n,1}$. Nevertheless,
amplitudes constructed by different deformations, such as
$M_n^{(1^\alpha,2_\alpha)}$ given by Eq.(\ref{Mn12}) can have
correct factorizations at (some of) these poles, then since we have
equalities relating them, they give the same amplitude as a rational
function of external momenta, which implies that
$M_n^{(1^\beta,n_\beta)}$ must also have the correct factorizations
at these poles.

If there are still some poles that are not explicitly included in
either Eq.(\ref{Mn1n}) or Eq.(\ref{Mn12}), then more deformations
which give the same function are needed. Therefore, our strategy
below is to find a chain of equalities which relates different
deformations to ensure each of them has the correct factorizations
at all poles, including holomorphic and anti-holomorphic collinear
limits.
\bigskip

Let us show the correct factorizations of the $n$ particle rational
function constructed by BCFW recursion relations for all helicity
configurations. First we discuss the case with $h_1=+$, then there
are four possibilities for $(h_n,h_1,h_2)$, which are
$(-,+,-)$,$(+,+,+)$,$(-,+,+)$ and $(+,+,-)$.

If $(h_n,h_1,h_2)=(-,+,-)$, then our proof in \ref{proof1} gives
$M_n^{(1^\alpha,2_\alpha)}=M_n^{(1^\beta,n_\beta)}$. But we can also
move a step towards the particle $n-1$, i.e., for any $h_{n-1}$, we
always have $M_n^{((n-1)^\mu,n_\mu)}=M_n^{(1^\beta,n_\beta)}$ since
$h_n=-$. Similarly, for any $h_3$, we have
$M_n^{(3^\nu,2_\nu)}=M_n^{(1^\alpha,2_\alpha)}$ since $h_2=-$. We
thus conclude
\begin{equation}
M_n^{((n-1)^\mu,n_\mu)}=M_n^{(3^\nu,2_\nu)}.
\end{equation}

Now we can check their factorizations at various poles. We have
shown that $M_n^{((n-1)^\mu,n_\mu)}$ has correct factorizations at
all possible poles except the holomorphic pole from $s_{n-2,n-1}$,
the anti-holomorphic pole from $s_{n,1}$, and both poles from
$s_{n-1,n}$. For $M_n^{(3^\nu,2_\nu)}$, we have shown it has the
correct factorizations at all possible poles except the holomorphic
pole from $s_{3,4}$, anti-holomorphic pole from $s_{1,2}$ and both
poles from $s_{2,3}$.

Since they are the same function, $M_n^{((n-1)^\mu,n_\mu)}$ must
have the same factorizations as $M_n^{((3)^\nu,2_\nu)}$, and vise
versa. The conclusion is that each of them does contain all possible
poles of an $n$ particle partial amplitude and has the correct
factorizations at each of them, which means that either by
$M_n^{((n-1)^\mu,n_\mu)}$ or $M_n^{(3^\nu,2_\nu)}$, or the same
function given by other deformations, we have obtained the correct
amplitude.

If $(h_n,h_1,h_2)=(+,+,+)$, then we have
$M_n^{(2^{\alpha'},1_{\alpha'})}=M_n^{(n^{\beta'},1_{\beta'})}$,
$M_n^{(n^{\mu'},(n-1)_{\mu'})}=M_n^{(n^{\beta'},1_{\beta'})}$, and
$M_n^{(2^{\nu'},3_{\nu'})}=M_n^{(2^{\alpha'},1_{\alpha'})}$.
Therefore, we arrive at
\begin{equation}
M_n^{(n^{\mu'},(n-1)_{\mu'})}=M_n^{(2^{\nu'},3_{\nu'})},
\end{equation}
where the L.H.S. explicitly has the correct factorizations at all
possible poles except the holomorphic pole from $s_{n,1}$,
anti-holomorphic pole from $s_{n-2,n-1}$ and both poles from
$s_{n-1,n}$, so does the R.H.S. at all possible poles except the
holomorphic pole from $s_{1,2}$, anti-holomorphic pole from
$s_{3,4}$, and both poles from $s_{2,3}$. Just as in the case
$(h_n,h_1,h_2)=(-,+,-)$, both of them have correctly included all
possible poles and yield the correct partial amplitude.

For $(h_n,h_1,h_2)=(-,+,+)$, we can only get a shorter chain of
equalities, i.e.,
\begin{equation}
M_n^{(1^\alpha,2_\alpha)}=M_n^{(1^\beta,n_\beta)}=M_n^{((n-1)^{\mu},n_{\mu})},
\end{equation}
for any $h_{n-1}$. This is because if we want to extend it to the
deformation on $(3,2)$, we may encounter the bad deformation if $h_3=-$.
However, we now show this shorter chain is enough for our purpose.

We have shown that $M_n^{(1^\alpha,2_\alpha)}$ explicitly has the
correct factorizations at all possible poles except the holomorphic
pole from $s_{n,1}$, anti-holomorphic pole from $s_{2,3}$, and both
poles from $s_{1,2}$, so does $M_n^{(1^\beta,n_\beta)}$ at all
possible poles except the holomorphic pole from $s_{1,2}$, anti-holomorphic pole from $s_{n-1,n}$, and both poles from $s_{n,1}$.

Since they are the same rational function, both
$M_n^{(1^\alpha,2_\alpha)}$ and $M_n^{(1^\beta,n_\beta)}$ have the
correct factorizations at all poles except the holomorphic poles
from $s_{n,1}$ and $s_{1,2}$.

Now since they are also the same function as
$M_n^{((n-1)^{\mu},n_{\mu})}$ which has the correct factorizations
at both poles from $s_{1,2}$ and the holomorphic pole from
$s_{n,1}$, we can see that all factorization channels of a partial
amplitude are correctly included and any of these deformations has
given the correct answer. The case with $(h_n,h_1,h_2)=(+,+,-)$ can
also be similarly proved by the chain
$M_n^{(1^\beta,n_\beta)}=M_n^{(1^\alpha,2_\alpha)}=M_n^{(3^\nu,2_\nu)}$.

All these discussions can apply to $h_1=-$, only with all helicities
flipped and $\lambda\leftrightarrow\tilde{\lambda}$. To summarize,
we have proved the weak version for $n$ particle partial amplitude,
namely any $n$ particle partial amplitude can be consistently
constructed from lower amplitudes by the deformation on any pair of
adjacent particles as long as it is a good deformation, and the
resultant function possesses all the correct factorization channels.
\bigskip

By induction, the conclusion is that for spin $1$ massless particles
in four dimensional Minkowski spacetime, given Poincare symmetry,
any tree-level amplitude can be constructed consistently from lower
amplitudes and eventually from basic three particle amplitudes via
BCFW recursion relations, if and only if, (1).\ there is no coupling
constants with negative dimensions, i.e.,
$\kappa^{[-2]}=\kappa'^{[-2]}=0$; and (2).\ dimensionless coupling
constants must conform to Jacobi condition, i.e.,
$\sum_{e}(f_{ade}f_{ebc}+f_{ace}f_{edb}+f_{abe}f_{ecd})=0$.
\section{Conclusion and Discussions}
In this paper we have investigated the consistency conditions on
scattering amplitudes of spin $1$ massless particles purely from the
S-matrix arguments. Instead of using Yang-Mills Lagrangian and its
Feynman diagrams, we directly constructed tree-level amplitudes from
lower amplitudes by BCFW recursion relations and proved this can be
consistently done and the resultant functions are indeed the correct
physical amplitudes. The main conclusions of this paper and
~\cite{BC} can be summarized as follows:\\
(1).\ Candidates for $n$ particle amplitudes are constructed from
lower amplitudes by BCFW recursion relations using a pair of
deformed particles with complex momenta.\\
(2).\ Three particle amplitudes are non-perturbatively determined by
Poincare symmetry and the tree-level four-particle test requires the
absence of negative dimension
couplings and dimensionless coupling constants to be the structure constants of a Lie group.\\
(3).\ Equalities relating candidates for the tree-level $n$ particle
amplitudes are obtained and are shown to have correct factorizations
at all possible poles, including holomorphic and anti-holomorphic
collinear limits, which ensure them to be the correct physical
amplitudes.

A remark on the strong version of consistency conditions is
necessary. We have only proved such equalities as
$M_n^{(i,i-1)}=M_n^{(i,i+1)}$ and $M_n^{(i-1,i)}=M_n^{(i+1,i)}$,
which are enough to ensure any of these deformations, as long as it
is a good one, yields the correct physical amplitude. This in turn
gives us a single stronger chain of equalities,
\begin{equation}
M_n^{(i,i+1)}=M_n^{(j,j+1)}=M_n^{(i,i-1)}=M_n^{(j,j-1)},
\end{equation}
for any $1\leq i,j\leq n$ as long as the concerned deformations are
all good.

What has not been proved is whether this equals any other good
deformation on non-adjacent particles, and a direct comparison of
amplitudes constructed by deforming non-adjacent particles with
those constructed by deforming adjacent particles, is at least not
very straightforward, due to the explicit use of color-decomposition.

A strategy to prove the strong version is to use the fact that by
any good deformation on adjacent particles we have obtain the
correct physical amplitude $M_n$, which in turn can be deformed on
any pair of non-adjacent particles, say $(l,m)$, as long as the
deformation is good. The key requirement for $M_n$ to be constructed
from lower amplitudes by deforming $(l,m)$ is
\begin{equation}
\lim_{z\rightarrow\infty}M_n^{l,m}(z)=0,
\end{equation}
which ensures $M_n^{(l,m)}\equiv M_n^{(l,m)}(0)$ to be expressed as
the sum of residues at finite poles and the result is exactly BCFW
construction.

Let us assume the strong version of consistency conditions on lower
amplitudes, which means any lower amplitude vanishes when a pair of
particles are deformed with parameter $z$ going to infinity. Now
since $n$ particle amplitude has been proven to be given by
$M_n^{(i,i+1)}$, if we deform a pair of particles $(l,m)$ of it and
send the parameter $z$ to infinity, it should be possible to show
that every term appearing in $M_n^{(i,i+1)}$ vanishes because lower
amplitudes vanish in this limit, from which the strong version of
consistency conditions can follow by induction.
\bigskip

There are several future directions worthy of investigations. An
obvious one is to generalize the proof of consistency conditions to
theories of particles with other spins, such as theories of spin $2$
massless particles and theories of particles with lower spins
coupled to spin $1$ or spin $2$ particles. Although the lack of
color-decomposition makes such generalizations apparently difficult,
it has been pointed out in~\cite{AK,ACK} that theories with spin $2$
particles, such as General Relativity and Supergravity, which do not
possess color-decomposition, have simpler structure in their
amplitudes due to even better vanishing behaviors at infinite
momenta, thus a similar proof of consistency conditions on
amplitudes in these gravitational theories is highly desirable.
Supersymmetric theories are notable here since supersymmetry can
relate amplitudes of particles with lower spins to the better
behaved amplitudes of highest spin particle, as have been used in
supersymmetric extension of BCFW in~\cite{ACK}. It will be
intriguing to see this purely from the S-matrix arguments.

In addition, it is interesting to see if this proof can be
generalized to other spacetime dimensions. Since BCFW recursion
relations have been proved for $D\geq 4$ dimensional Yang-Mills
theory and perturbative gravity from Lagrangian point of
view~\cite{AK}, there should be a direct generalization of our proof
to higher dimensions although the convenient spinor techniques may
not be used in this case. A crucial insight is the enhanced Lorentz
symmetry of effective theory at large complex momenta and it is
desirable to uncover it without Lagrangian or Feynman diagrams. On
the other hand, consistency conditions on theories in three
dimensions are also interesting since exactly solvable models are
available there.

A direction of more significance is the investigation of purely
S-matrix argument for simplicities of loop-level amplitudes and
their consistency conditions. From quantum field theory point of
view, tree-level consistent theories can be anomalous at
loop-levels, thus it is important to extend our analysis to the
loop-levels. On the other hand, remarkable simplicities in
loop-level amplitudes which are not manifest from local quantum
field theory and its Feynman diagrams, especially for maximal
supersymmetric theories in four dimensional spacetime, i.e.,
$\mathcal{N}=4$ SYM theory and $\mathcal{N}=8$ Supergravity, imply
that a purely S-matrix understanding is necessary and desirable. For
example, it will be interesting to derive the absence of triangles,
bubbles and rational terms at one-loop level for maximal
supersymmetric theories from a purely S-matrix argument.

An even more ambitious possibility, as emphasized in~\cite{ACK}, is
to search for a dual formulation of local quantum field theory and
its Feynman diagrams which manifests BCFW construction as well as
loop-level simplicities of amplitudes. Since now a purely S-matrix
argument for consistency conditions on amplitudes is available, the
existence of such a dual formulation has been put onto a more solid
ground and the proof here can be considered as a starting point for
the construction of the dual theory.

\section*{Acknowledgement}
We are grateful to Freddy Cachazo and Huaxin Zhu for stimulating
their interests in this topic. We also thank Muxin Han and Yidun Wan
for interesting discussions. SH is much indebted to Freddy Cachazo
for many enlightening discussions which make this work possible, and
for important comments on the draft. He also thanks Bo Feng and
Huaxin Zhu for valuable discussions and comments on the draft. SH
was supported by NFSC grants(no.10235040 and 10421003). HZ was
supported in part by the Government of China through
CSC(no.2007102530). This research was supported by Perimeter
Institute for Theoretical Physics. Research at Perimeter Institute
is supported by the Government of Canada through IC and by the
Province of Ontario through MRI.

\section*{Appendix: Expressions of deformed momenta}
\begin{appendix}\setcounter{equation}{0}
Here we shall work out the specific expressions for
$1^\alpha,2_\alpha,1^\beta,n_\beta,1^\beta\oplus2$ and
$n\oplus1^\alpha$. First $1^\alpha$ and $2_\alpha$ come from the
deformation $(1^\alpha,2_\alpha)$, i.e.,
\begin{eqnarray}\label{alpha}
\lambda^{(1^\alpha)}=\lambda^{(1)}+z(\alpha)\lambda^{(2)},\tilde{\lambda}^{(1^\alpha)}=\tilde{\lambda}^{(1)},\nonumber\\
\lambda^{(2_\alpha)}=\lambda^{(2)},\tilde{\lambda}^{(2_\alpha)}=\tilde{\lambda}^{(2)}-z(\alpha)\tilde{\lambda}^{(1)},
\end{eqnarray}
whereby the parameter $z(\alpha)$ can be obtained as
$z(\alpha)=-\langle n,1\rangle/\langle n,2\rangle$ by leaving
on-shell the momentum of $n\oplus 1^\alpha$,
\begin{equation}
P=\lambda^{(1^\alpha)}\tilde{\lambda}^{(1)}+\lambda^{(n)}\tilde{\lambda}^{(n)}.
\end{equation}
Furthermore taking the inner product of the momentum with the
$\lambda^{(n)}$ yields zero by using $\langle n,1^\alpha\rangle=0$.
Whence we know the left-handed part can always be set equal to
$\lambda^{(n)}$ due to the little group transformation. Thus by
taking the inner product of the momentum with $\lambda^{(1)}$, we
obtain the final expression of $n\oplus1^\alpha$ as
\begin{equation}\label{noplus1alpha}
\lambda^{(n\oplus1^\alpha)}=\lambda^{(n)},\tilde{\lambda}^{(n\oplus1^\alpha)}=\frac{\langle1,2\rangle}{\langle
n,2\rangle}\tilde{\lambda}^{(1)}+\tilde{\lambda}^{(n)}.
\end{equation}

Similarly, $1^\beta$ and $n_\beta$ come from the deformation
$(1^\beta,n_\beta)$, i.e.,
\begin{eqnarray}\label{beta}
\lambda^{(1^\beta)}=\lambda^{(1)}+z(\beta)\lambda^{(n)},\tilde{\lambda}^{(1^\beta)}=\tilde{\lambda}^{(1)},\nonumber\\
\lambda^{(n_\beta)}=\lambda^{(n)},\tilde{\lambda}^{(n_\beta)}=\tilde{\lambda}^{(n)}-z(\beta)\tilde{\lambda}^{(1)},
\end{eqnarray}
whereby the parameter $z(\beta)$ can be obtained as
$z(\beta)=-\langle1,2\rangle/\langle n,2\rangle$ by leaving on-shell
the momentum of $1^\beta\oplus2$,
\begin{equation}
P=\lambda^{(1^\beta)}\tilde{\lambda}^{(1)}+\lambda^{(2)}\tilde{\lambda}^{(2)}.
\end{equation}
Furthermore taking the inner product of the momentum with the
$\lambda^{(2)}$ yields zero by using $\langle1^\beta,2\rangle=0$.
Whence we know the left-handed part can always be set equal to
$\lambda^{(2)}$ due to the little group transformation. Thus by
taking the inner product of the momentum with $\lambda^{(1)}$, we
obtain the final expression of $1^\beta\oplus2$ as
\begin{equation}\label{1betaoplus2}
\lambda^{(1^\beta\oplus2)}=\lambda^{(2)},\tilde{\lambda}^{(1^\beta\oplus2)}=\frac{\langle
n,1\rangle}{\langle
n,2\rangle}\tilde{\lambda}^{(1)}+\tilde{\lambda}^{(2)}.
\end{equation}
\end{appendix}

\end{document}